# Multi-pole multi-zero frequency-independent phase-shifter

M.A. Bitar,[1] A. Gallo,[2] and F.A. Volpe[3,a)]

[1]*American University of Beirut, Beirut, Lebanon*
[2]*INFN, Laboratori Nazionali di Frascati, 00044 Frascati (Roma), Italy*
[3]*Dept of Applied Physics and Applied Mathematics, Columbia University, New York, NY 10027, USA*

A multi-pole, multi-zero design allowed realizing a "true" phase-shifter (not time-delayer) of flat frequency-response over more than 3 decades (30Hz-100kHz), which can be extended to higher frequencies or broader bands thanks to a modular design. Frequency-dependent optimization of a single resistance made also the gain flat to within few percents. The frequency-independent phase-shifter presented can find application in any experiment in which an action needs to be taken (e.g. a measurement needs to be performed) at a fixed phase-delay relative to an event, regardless of how rapidly the system rotates or oscillates.

## I. INTRODUCTION

Most phase-shifters are in fact time-delayers introducing a nearly frequency-independent time-delay $\Delta t$, proportional to the time-constant of the circuit. "True" phase-shifters introducing a frequency-independent phase-delay $\Delta \phi$ are relatively rare, despite numerous potential applications. Such applications regard any rotary or oscillatory system (from particle accelerators[1] to fusion plasmas[2,3], from turbines to internal combustion engines, pumps and lock-in amplifiers) in which an operation (for instance deploying an actuator, or performing a measurement) has to be performed at a fixed *phase* delay with respect to an event (such as a sensor measurement or a perturbation) regardless of how rapidly the system rotates or oscillates.

This was achieved in the past by means of a resistance $R$ varying with the angular frequency $\omega$ in such a way that the time-constant of the circuit, $\tau = RC$, decayed like $1/\omega$ and thus the phase shift $\omega\tau$ remained constant. The desired dependence of $R$ on $\omega$ was obtained by combining a Frequency-to-Voltage Converter (FVC) with a Voltage-controlled Resistor (VCR) such as a matched pair of FETs[1,4] or a digital potentiometer interfaced to a micro-controller[2,3]. These circuits operated at, respectively, 20-40kHz [4], 1-7kHz [2,3] and up to 1-20MHz [1]. The last one, in particular, covered more than one decade.

Here we present a simple analog circuit yielding flat response over more than 3 decades ($f = $ 30Hz-100kHz). The relative bandwidth can be made even broader thanks to a modular design: the circuit is a series of "stages", each imparting a different, frequency-dependent phase-shift (Fig.1). The idea is that for a proper choice of zeros and poles, the phase-shifts peak at different frequencies in different stages and their series yields an approximately flat response (Fig.1). Secs.2 and 3 describe analytical and numerical methods to identify the optimal poles and zeros enabling this result. These translate in constraints on the resistors and capacitors to be used. A 2-stage and a 5-stage shifter were constructed accordingly and were verified to exert the expected phase-shift over the expected broad band (Sec.4).

a) Author to whom correspondence should be addressed:
fvolpe@columbia.edu.

## II. ANALYTHICAL OPTIMIZATION FOR 2 STAGES

Consider the series of *n* stages, each characterized by a pole $\omega_{pi}$ and a zero $\omega_{zi}$, where $i$=1,2,…,*n*. An example for *n*=5 is shown in Fig.2, with an additional inverting amplifier to compensate for the 180° phase-shift due to the odd number of stages. Another circuit examined, not shown for brevity, had *n*=2 stages and no inverting amplifier. The transfer function is the product of the transfer functions for each stage:

$$H = (-1)^n \prod_{i=1}^n A_{0i} \frac{1+s/\omega_{zi}}{1+s/\omega_{pi}}, \quad (1)$$

where $A_{0i}$ are the d.c. gain and $s = \sigma + i\omega$ the Laplace transform parameter. At angular frequency $\omega$, the circuit introduces a phase-shift

$$\Delta\varphi(\omega) = \sum_{i=1}^n \left(\arctan\frac{\omega}{\omega_{zi}} - \arctan\frac{\omega}{\omega_{pi}}\right), \quad (2)$$

exhibiting ripples in the $[\omega_{c1}, \omega_{cn}]$ band, where $\omega_{ci} = \sqrt{\omega_{zi}\omega_{pi}} = \omega_{zi}/\sqrt{a_i} = \sqrt{a_i}\omega_{pi}$ is the central frequency for stage *i*. Note that transcendental Eq.2 can be casted in rational form by repetitive applications of the identity [5]

$$\arctan z_1 \pm \arctan z_2 = \arctan\left(\frac{z_1 \pm z_2}{1 \mp z_1 z_2}\right). \quad (3)$$

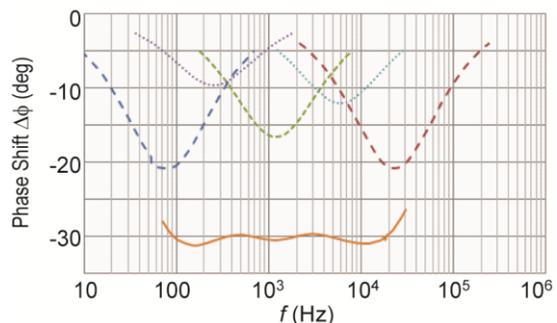

FIG. 1. (Color online). Calculated Bode plots of (a) gain and (b) phase-shift imparted by each stage, and their cumulative effect, for the five-stage design in Fig.2, using ideal components.

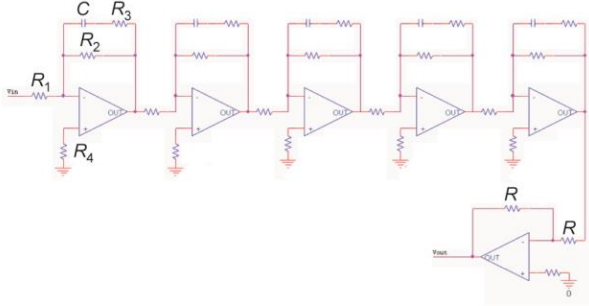

FIG. 2. (Color online). Five-stage phase-shifter. For simplicity, only the components of the first stage have been labeled.

With this, let's consider $n=2$ stages and, in order for the ripple to be small, let's impose the same phase-shift $\Delta\varphi(\omega) = \theta$ at $\omega = \omega_{c1}$, $\omega = \omega_{c2}$ and at their geometric average:

$$\tan\theta \cdot \left(1 + u^2 - \frac{uvw}{2}\right) = (1 + u^2)\frac{v}{2} + uw \quad (4)$$

$$\tan\theta \cdot \left(1 + \frac{1}{u^2} - \frac{vw}{2u}\right) = \left(1 + \frac{1}{u^2}\right)\frac{w}{2} + \frac{v}{u} \quad (5)$$

$$\tan\theta \cdot \left(\frac{(1+u)^2}{u} - vw\right) = \left(\sqrt{u} + \frac{1}{\sqrt{u}}\right) \cdot (v + w) \quad (6)$$

Here $u = \omega_{c,1}/\omega_{c2}$ is a multiplier relating the central frequencies to each other. $v = (1 - a_1)/\sqrt{a_1}$ and $w = (1 - a_2)/\sqrt{a_2}$ are also combination of multipliers, $a_i = \omega_{zi}/\omega_{pi}$.

From Eqs.5-6 we can conclude that $w = v$, hence $a_1 = a_2 = a$. After these substitutions, Eq.6 returns

$$v = C\frac{1+u}{\sqrt{u}} \quad (7)$$

where the coefficient $C = \frac{(-1 \pm \sqrt{1+(\tan\theta)^2})}{\tan\theta}$ is fixed by the desired phase-shift $\theta$. By eliminating $v$ in Eq.4, we obtain the following algebraic equation in the unknown $x = \sqrt{u}$:

$$x^6 + \left(\frac{C^2-2}{C}\tan\theta\right)x^5 + 3x^4 + (2C\tan\theta)x^3 + 3x^2 + \left(\frac{C^2-2}{C}\tan\theta\right)x + 1 = 0. \quad (8)$$

Solving this equation for $x$ leads to possible values of the multiplier $u = x^2$ relating $\omega_{c2}$ to $\omega_{c1}$. Then, choosing one central frequency indirectly fixes the other. Finally $u$ leads to $v$ (through Eq.7) and ultimately to $a$.

At this point we have determined the two central frequencies $\omega_{ci}$ and the (identical) multipliers $a_i$. As a result we have fixed the two poles and zeros $\omega_{pi}$ and $\omega_{zi}$; resistors and capacitors need to be chosen accordingly. Note that while the set of optimal $\omega_{pi}$ and $\omega_{zi}$ leading to the desired phase-shift (Eq.4-6) is unique, the corresponding set of resistors and capacitors is not, which adds flexibility to the design. For the first stage in Fig.2, for example, $\omega_z = 1/CR_3$ and $\omega_p = 1/C(R_2 + R_3)$.

Note also that only one value of $C$ is of interest, namely $C = \frac{(-1+\sqrt{1+(\tan\theta)^2})}{\tan\theta}$. For this, Eq.8 admits two real solutions and four complex ones, whereas for the other value of $C$ it only admits complex solutions. Of the real roots we choose the value of $u$ which is less than 1, since we already mentioned that $u = \omega_{c1}/\omega_{c2}$, and here assume $\omega_{c1} < \omega_{c2}$.

Finally, there are always two solutions, $a < 1$ and $a > 1$. Depending on the desired phase-shift being positive or negative, one will adopt one or the other.

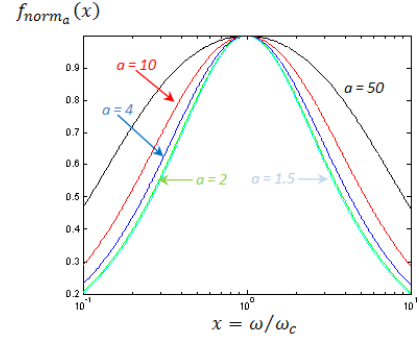

FIG. 3. (Color online). $f_{norm_a}(x)$ (Eq.17) as a function of $x = \omega/\omega_c$, for several values of $a$.

## A. Extension to arbitrary phase-shifts

The method can be applied to the search for solutions for $\theta \geq 30°$ for $n = 2$ stages shifters. For smaller phase-shifts, one can simply rescale the $\theta = 30°$ solution. The reason is illustrated in Fig.3: the phase response normalized to its peak value gives a "form factor" of the phase frequency response of each stage,

$$f_{norm_a}(x) = \frac{\arctan(x/\sqrt{a}) - \arctan(x\sqrt{a})}{\arctan(1/\sqrt{a}) - \arctan(\sqrt{a})} \xrightarrow{a\to 1} \frac{2x}{1+x^2}, \quad (9)$$

where $x = \omega/\omega_c$. For $a \leq 3$ (and thus, after Eq.5, $\theta \leq 30°$) the form factor does not change noticeably in Fig.9. This means that $\varphi(f)$ does not change significantly in shape with $\theta$, provided $\theta \leq 30°$. In turn, this implies that the solution $\varphi(f)$ (basically the optimal $u$ and $v$, or $a$) for $\theta = 30°$ is readily generalized to $\theta < 30°$ by a simple rescaling.

Finally, the present paper concentrated on phase-shifters of fixed phase shift $\Delta\varphi$, but adjustable $\Delta\varphi$ can be easily obtained by combining unshifted signals and 90°-shifted signals in an additive mixer, with the weights of the linear combination determining the phase-shift $\Delta\varphi$ as follows:

$$S\sin\omega t + C\cos\omega t = \sqrt{S^2 + C^2}\sin(\omega t + \Delta\varphi) \quad (10)$$

$$\Delta\varphi = \arctan\frac{C}{S} + \begin{cases} 0 \text{ if } S \geq 0 \\ \pi \text{ if } S < 0 \end{cases} \quad (11)$$

## III. NUMERICAL OPTIMIZATION FOR *n* STAGES

### A. Simplified Treatment

Let us assume for simplicity that ratios $a_i = \omega_{zi}/\omega_{pi}$ are the same for all stages. Let us also assume that each central frequency $\omega_{ci}$ is related to the central frequency of the next stage by the same multiplier $m = \omega_{c,i+1}/\omega_{ci}$ and let us define the phase non-uniformity as the standard deviation of $\Delta\varphi$ normalized to the mean of $\Delta\varphi$ over the range $[\omega_{p1}, \omega_{zn}]$, i.e. from the first pole to the last zero. The phase is flattest when this normalized standard deviation is smallest. Fig.4a shows this quantity as a function of $m$. The plot was obtained for $a_i = a = 2.433$ for all $i$.

It is intuitive that the best compromise of flatness and bandwidth is obtained for multipliers $m$ that are neither too small (otherwise the peaks are too close to each other and the band of high flatness is narrow) nor too large, otherwise the peaks are spaced by more than their widths, resulting in significant ripples. Fig.4 indicates that $m \approx 12$ is a good compromise for $a=2.433$. Other values of $a$ lead to similar plots, with minima at comparable locations, $m \gtrsim a^2$.

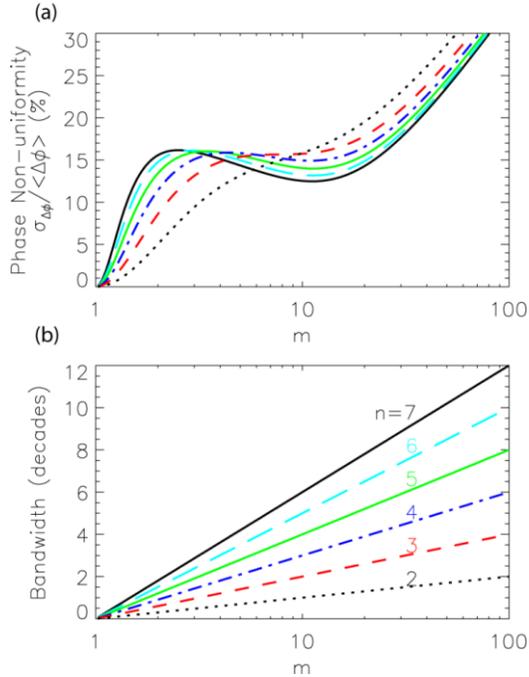

FIG. 4. (Color online). (a) Phase non-uniformity (defined as standard deviation of $\Delta\phi$, divided by its mean over the band of interest) and (b) bandwidth (defined as $(\omega_{cn}-\omega_{c1})/2\pi$) of an $n$ stage shifter, as a function of the multiplier $m = \omega_{c,j+1}/\omega_{cj}$.

Note that the choice of $a$ governs the steepness of $\Delta\varphi(\omega)$ outside of $[\omega_{c1}, \omega_{cn}]$, with lower $a$ yielding steeper functions. The number of stages $n$ determines the bandwidth (Fig.4b).

## B. General Treatment

The previous simplified treatment reached phase non-uniformities of 12-15% (Fig.4a). Flatter response can be achieved if the simplifying assumptions on $a_i$ and $m$ are removed. In this case the non-uniformity minimization problem, one-dimensional in Fig.4, becomes multi-dimensional, but its dimensionality is reduced by symmetry arguments: the center frequencies of the stages have to be distributed symmetrically in linear or logarithmic scale around the central frequency of the device. As a result, the total number of degrees of freedom is $n$. This is due to the number of independent frequencies being $n/2$ if $n$ is even, or $(n-1)/2$ if $n$ is odd. Likewise, symmetric stages have to introduce the same phase insertion, $\theta_i = \theta_{n+1-i}$. This reduces the number of variables $\theta_i$ (or equivalently $a_i$, through Eq.5) to $n/2$ if $n$ is even, or $(n+1)/2$ if $n$ is odd. In conclusion the problem is $n$-dimensional: $\Delta\varphi(\omega)$ is parameterized in $n$ parameters $a_i$ and $\omega_{ci}$. Constraints can be imposed to set the $[\omega_{c1}, \omega_{cn}]$ band of interest.

Let us define $\omega_c$ the central frequency and $a$ the relative bandwidth, meaning that the phase-shifter covers frequencies from $\omega_c/\sqrt{a}$ to $\omega_c\sqrt{a}$. Let us also introduce $k_j = \omega_{c_j}/\omega_c$ and $x = \omega/\omega_c$. From Eqs.10, 11 and symmetry arguments it follows that for $n$ even

$$\Delta\varphi(x) = \sum_{j=1}^{n/2} \arctan \frac{k_j\sqrt{a_j}(1-a_j)(1+k_j^2)(1+x^2)x}{a_j(k_j^2+x^2)(1+k_j^2x^2)-(1-a_j)^2 k_j^2 x^2}, \quad (12)$$

and for $n$ odd

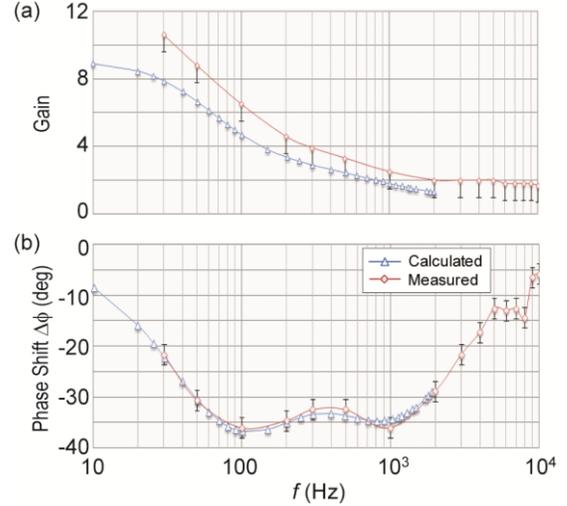

FIG. 5. (Color online). Bode plots for two-stage amplifier applying negative phase-shift $\theta=-34\pm 4°$ in the range $f$=55-2000Hz.

$$\Delta\varphi(x) = \arctan \frac{1}{\sqrt{a_0}} \frac{(1-a_0)x}{1+x^2} + \sum_{j=1}^{(n-1)/2} \arctan \frac{k_j\sqrt{a_j}(1-a_j)(1+k_j^2)(1+x^2)x}{a_j(k_j^2+x^2)(1+k_j^2x^2)-(1-a_j)^2 k_j^2 x^2}. \quad (13)$$

In both cases, there are $n$ free parameters $a_j$ and $\omega_{cj}$. One constraint comes from imposing that the frequency-averaged phase-shift $\Delta\varphi$ equals the desired value $\theta$:

$$\langle\Delta\varphi(x)\rangle = \frac{\sqrt{a}}{a-1}\int_{1/\sqrt{a}}^{\sqrt{a}}\Delta\varphi(x)dx = \theta. \quad (14)$$

The remaining $n$-1 degrees of freedom will be used to minimize the standard deviation

$$\sigma_{\Delta\varphi}^2 = \langle(\Delta\varphi)^2\rangle - \langle\Delta\varphi\rangle^2 = \frac{\sqrt{a}}{a-1}\int_{1/\sqrt{a}}^{\sqrt{a}}(\Delta\varphi)^2(x)dx - \theta^2. \quad (15)$$

Finally, in order for $\Delta\varphi$ to be flat in the Bode diagram, it is necessary to introduce a new coordinate $y = \log_{10} x$ and modify the last two expressions as follows:

$$\langle\Delta\varphi(y)\rangle = \frac{1}{\log_{10} a}\int_{-\frac{\log_{10} a}{2}}^{\frac{\log_{10} a}{2}} \Delta\varphi(y)dy = \theta, \quad (16)$$

$$\sigma_{\Delta\varphi}^2 = \frac{1}{\log_{10} a}\int_{-\frac{\log_{10} a}{2}}^{\frac{\log_{10} a}{2}} \theta^2(y)dy - \theta^2. \quad (17)$$

## IV. RESULTS FOR TWO AND FIVE STAGES

Following the procedure of Sec.2, the optimal zeros (1.94 and 0.16kHz) and poles (0.69 and 0.05kHz) were identified for an $n$=2 phase-shifter with the goal of flat response in the $\omega_{c2} < \omega < \omega_{c1}$ range and beyond. Here $\omega_{c1}/2\pi = 1154$Hz and $\omega_{c2}/2\pi = 89$Hz. The result was a flat response $\theta = -34 \pm 4°$ in the range $f = 55$-2000Hz (Fig.5b). The overall gain is the product of the gains, both decreasing with $f$, hence it also decreases with $f$ (Fig.5a).

The $n$=5 circuit in Fig.2 had the goal of maintaining the phase shift constant ($\theta = -30°$, equivalent to $\theta = 330°$) over an even broader range: $f = 0.1$-20kHz, as needed for the control of rotating instabilities in fusion plasmas [2,3].

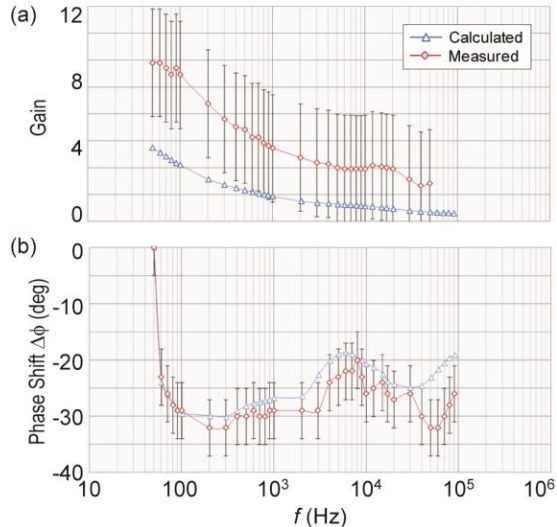

FIG. 6. (Color online). Calculated and measured Bode plots of cumulative (a) gain and (b) phase-shift for the five-stage design in Fig.2, using realistic components.

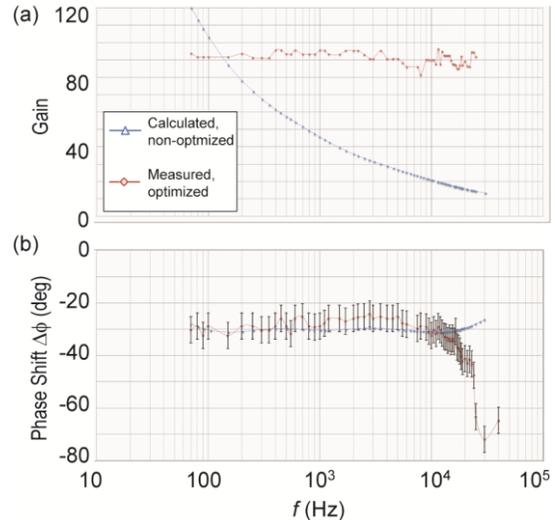

FIG. 7. (Color online). Measured (a) gain and (b) phase-shift introduced by a five-stage phase-shifter of adjustable gain, where $R_1$ of stage 1 (Fig.2) was optimized at every $f$ for flat gain. Both flat and phase are flat over a band broader than 0.07-10kHz. Calculations at fixed $R_1$ are also shown for comparison.

Its zeros, poles and ideal components were chosen as per the method of Sec.3B, although the actual resistances and capacitances were used to calculate the Bode plot in Fig.6. The measurements, in very good agreement with the calculations, exhibit a phase shift as flat as $\theta = 24\pm7^\circ$ over a band as wide as $f = 0.03$-100kHz (Fig.6b).

The gain decays with $f$ as expected (Fig.6a). In some cases, however, in addition to constant phase-shift it is useful to also have constant gain along the frequency band of interest. For this purpose, a variant of Fig.2 was realized by simply replacing the resistor $R_1$ in stage 1 with a variable resistor. For every $f$, the variable resistor was manually adjusted so that the gain remained constant. The measurements are presented in Fig.7. The adjustment of $R_1$ as a function of $f$ was performed by hand, but it is conceivable that replacing $R_1$ with an FVC and VCR should allow automating such adjustment, and therefore simultaneously achieving flat response in phase and gain, similar to Fig.7 but with no need for manual adjustment. Note that a single resistance ($R_1$) in a single stage needs to be adjusted. Note also that the flatness of the gain is decoupled from the flatness of the phase-shift, which, through poles and zeros, only depends on $R_2$, $R_3$ and $C$ in the various stages, but not on $R_1$ (see Fig.2).

## V. CONCLUSIONS

In summary simple modules featuring a single pole and single zero were cascaded so that the phase shifts add up and their frequency dependencies compensate each other. This resulted in the frequency response of the *phase* shift (not the *time* delay) being flat to within few percents over more than 3 decades. Gain adjustment allowed the gain to be comparably flat over the same band. The circuit might find application whenever an operation needs to be performed at a fixed *phase* delay with respect to an event, regardless of how rapidly or slowly the system rotates or oscillates. The concept is modular and easily extendible to other and/or broader bands, thanks to a simple optimization method.